\documentclass[prd,aps,twocolumn,showpacs,nofootinbib]{revtex4-1}

\usepackage{color}

\definecolor{green}{rgb}{0,.5,0}

\definecolor{red}{rgb}{1,0,0}

\usepackage{graphicx}
\usepackage[subfigure]{graphfig}
\usepackage{epsfig}
\usepackage{dcolumn}
\usepackage{amsmath,amssymb,amsthm}
\usepackage{latexsym}
\usepackage{multirow}
\usepackage{bbm}
\usepackage{ulem}
\def\be{\begin{equation}}
\def\ee{\end{equation}}
\def\bea{\begin{eqnarray}}
\def\eea{\end{eqnarray}}

\begin{document}

\title{Gluons in charmoniumlike states}

\author{Wei Sun$^{1}$, Ying Chen$^{1,2}$, Peng Sun$^{3}$, Yi-Bo Yang$^{4,5,6}$
\vspace*{-0.5cm}
\begin{center}
\large{
\vspace*{0.4cm}
\includegraphics[scale=0.15]{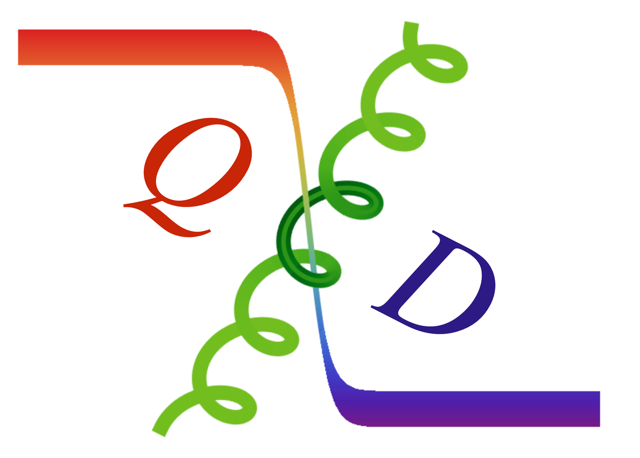}\\
\vspace*{0.4cm}
($\chi$QCD Collaboration)
}
\end{center}
}
\affiliation{
$^{1}$\mbox{Institute of High Energy Physics, Chinese Academy of Sciences, Beijing 100049, China}\\
$^{2}$\mbox{School of Physics, University of Chinese Academy of Sciences, Beijing 100049, China}\\
$^{3}$\mbox{Nanjing Normal University, Nanjing, Jiangsu, 210023, China}\\
$^{4}$\mbox{CAS Key Laboratory of Theoretical Physics, Institute of Theoretical Physics, Chinese Academy of Sciences, Beijing 100190, China}\\
$^{5}$\mbox{School of Fundamental Physics and Mathematical Sciences, Hangzhou Institute for Advanced Study, UCAS, Hangzhou 310024, China}\\
$^{6}$\mbox{International Centre for Theoretical Physics Asia-Pacific, Beijing/Hangzhou, China}\\
}

\begin{abstract}
	The mass components of charmoniumlike states are investigated through the decomposition of QCD energy-momentum tensor (EMT) on lattice.
	The quark mass contribution $\langle H_m\rangle$ and the momentum fraction $\langle x\rangle$ of valence charm quark
	and antiquark are calculated for conventional $1S,1P,1D$ charmonia and the exotic $1^{-+}$ charmoniumlike state,
	based on the $N_f=2+1$ gauge configurations generated by the RBC/UKQCD collaboration.
	It is found that $\langle H_m\rangle$ is close to each other and around 2.0 to 2.2 GeV for these states, which implies
	that the mass splittings among these states come almost from the gluon contribution of QCD trace anomaly. The $\langle x\rangle$ of the $1^{-+}$ state is only around 0.55,
	while that in conventional charmonia is around 0.7 to 0.8.
	This difference manifests that the proportion of light quarks and gluons
	in the $1^{-+}$ charmoniumlike state is significantly larger than conventional states.
\end{abstract}

\maketitle

\section{Introduction}\label{intro}
Based on the quantum chromodynamics (QCD) the gluon is massless and the intermedia of the strong interaction.
 It is well-know that gluons bind light quarks into massive hadrons, then there is a question that how much gluons
contribute to the total mass of a hadron. In order to answer this question, one can start from the hadron rest frame energy decomposition~\cite{Ji:1994av},
\bea
 M = \langle H_m\rangle + \langle H_E\rangle (\mu)+ \langle H_g\rangle (\mu) +  \frac{1}{4} \langle H_a \rangle, \label{eq:total}
\eea
where $H_m$, $H_E$, $H_g$ and $H_a$ denote the parts of the Hamiltonian contributed
by the quark mass, quark energy, glue energy and QCD anomaly, respectively, whose explicit forms are
\begin{align}\label{eq:decomp}
    H_{m} &=\sum_{q=u, d, s \cdots} \int d^{3} x m_q \bar{\psi}_q \psi_q,\nonumber\\
	H_{E} &=\sum_{q=u, d, s \ldots} \int d^{3} x \bar{\psi}_q(\vec{D} \cdot \vec{\gamma}) \psi_q,\nonumber\\
	H_{g} &=\int d^{3} x \frac{1}{2}\left(B^{2}-E^{2}\right),\nonumber\\
    H_{a} &=\int d^{3} x \big[\sum_{q=u, d, s \cdots}\gamma_{m} m_q \bar{\psi}_q \psi_q - \frac{\beta(g)}{g}\left(E^{2}+B^{2}\right)\big].
\end{align}
Here $\langle \cdots\rangle$ means $\langle h|\ldots|h\rangle/\langle h|h \rangle$ with $|h\rangle$
being the hadron state in its rest frame. Considering the trace sum rule $M= \langle H_m\rangle+\langle H_a \rangle$~\cite{Shifman:1978zn},
there are only two independent components in Eq.~(\ref{eq:total}). Besides that, there is another decomposition proposed in~\cite{Lorce:2017xzd}. Regardless how the decomposition is performed, the gluon contribution is always comparable with that from quarks for light hadrons~\cite{Yang:2018nqn,Yang:2014xsa}.

The spectrum of heavy quarkonia are usually studied by nonrelativistic quark model, 
where part of the effect of gluons is reflected through the confining potential. 
In order to investigate the role played by gluons in heavy quarkonia and heavy quarkoniumlike states 
from the point of view of EMT, we consider the mass decomposition of both the conventional charmonium states, 
such as $1S(\eta_c, J/\psi)$, $1P(h_c, \chi_{c0,1,2})$, $1D(\eta_{c2})$ charmonia, 
and the $1^{-+}$ charmoniumlike state in the lattice QCD formalism. 
The major goal is to check the contribution of each part of the Hamiltonian in Eq.~(\ref{eq:decomp}) 
to the masses of these states and their mass splittings. There is a special interest in the $1^{-+}$ charmoniumlike state, 
since its quantum number is prohibited by a $c\bar{c}$ system in the quark model picture, 
but is permitted either by a $c\bar{c}g$ hybrid or multiquark system involving a $c\bar{c}$ pair. 
The gluon contribution to its mass in comparison with that of the conventional charmonia may shed light 
on the nature of the $1^{-+}$ charmoniumlike state.  

The lattice calculation in this work is based on the overlap fermion and $N_f=2+1$ domain wall gauge configurations. 
On a finite Euclidean lattice, the energy levels of the Hamiltonian have a discrete spectrum of values, the connection 
between the lattice energy eigenstates and the physical hadron states are not usually straightforward, 
since most of hadrons appear as resonances.  
The masses of the $1S$ and $1P$ charmonia are below the $D\bar{D}$ threshold, such that their strong decays 
are suppressed by the Okubo-Zweig-Iizuka rule and result in their small decay widths. 
The expected $1D(\eta_{c2})$ mass is approximately 3.8 GeV~\cite{Liu:2012ze,Yang:2012mya} 
and likely lower than the $D\bar{D}^{*}$ threshold, which is the symmetry-permitted lowest open charm threshold. 
Thus the width of $\eta_{c2}$ can be very small. Therefore, the $1S, 1P, 1D$ charmonium states in this work 
can be viewed as stable particles and have direct correspondence to the related states on the lattice. 
However, for the $1^{-+}$ charmoniumlike state, even though it has not been observed in experiments, 
previous lattice calculations predict its mass to be around 4.3 GeV~\cite{Yang:2012gz,Liu:2012ze}, 
which is well above the open charm threshold. In a strict meaning, one must establish the connection 
of the lattice states in this channel to the possible physical states 
in the L\"{u}scher formalism~\cite{Luscher:1986pf,Luscher:1990ux} by studying the related meson-meson scatterings. 
This requires certainly sophisticated numerical techniques to tackle the annihilation diagrams of light quarks 
and to derive precise energy levels. As an exploratory study, we tentatively ignore the decay effects 
of the $1^{-+}$ charmoniumlike state and view it as single particle state in the data analysis of relevant two-point 
and three-point functions in this work, a more rigorous 
treatment is left for the future studies.

The paper is organized as follows: In Sec. \ref{method} we describe the lattice setup, construction of correlation
functions, numerical method used to extract the matrix elements of EMT, and the simultaneous fit strategy.
Section \ref{result} discusses the ground state mass, quark mass contribution $\langle H_m \rangle$,
valence charm quark momentum fraction $\langle x \rangle_q$ in $1S(\eta_c, J/\psi)$, $1P(h_c, \chi_{c0,1,2})$, $1D(\eta_{c2})$ and $1^{-+}$ states. 
Section \ref{summary} summarizes the main results of this paper.

\section{Numerical details}\label{method}
We perform the calculation using the 2+1 flavor configurations of domain wall fermion
and Iwasaki gauge action provided by RBC/UKQCD collaboration.
The parameters of two gauge ensembles~\cite{Aoki:2010dy,Mawhinney:2019cuc} we used are listed in Table~\ref{table:ensemble}.
The overlap fermion~\cite{Chiu:1998gp} is adopted for the valence charm quark, and we tune the bare quark mass parameters
on the two ensembles to reproduce the physical $J/\psi$ mass $M_{J/\psi}=3.097$ GeV within 0.2\%.
By using the overlap fermion, $\langle H_m\rangle$ is automatically renormalization scale and scheme independent,
and then the other components are well defined thanks to the quark equation of motion~\cite{Yang:2014xsa}.
We use the conventional quark bilinear operators $\bar{c}\Gamma(D)c$
for $1S(\eta_c,J/\psi)$, $1P(h_c,\chi_{c0,1,2}$) and $1D(\eta_{c2})$ charmonia, where $\Gamma$
stands for the specific combinations of Dirac gamma matrices and $D$ is the lattice covariant derivative operator.
For the $1^{-+}$ charmoniumlike state, the operator is chosen to be
$\epsilon_{ijk}\bar{c}\gamma_j B_k c$ where $B_k$ is the chromomagnetic strength tensor.
The details of the lattice interpolation operators and their quantum numbers can be found in Table~\ref{tab:mass_table},
where the available experiment results of masses of conventional charmonia in Ref.~\cite{Zyla:2020zbs} are also listed there.
\begin{table}[t]
	\caption{\label{table:ensemble}
	The parameters of gauge ensembles. The pion mass $m_\pi$ corresponds to the $u,d$ sea quark mass parameter,
	while the bare quark mass parameter $m_ca$ on each ensemble is set by the physical $J/\psi$ mass $M_{J/\psi}=3.097$ GeV,
	and $N_\text{cfg}$ is the number of configurations used in the calculation.}
	\begin{ruledtabular}
		\begin{tabular}{c c c c c c}
			\text{ensemble} & $L^3 \times T$   &  $a$ (fm)   & $m_\pi$ (MeV)  &  $m_ca$   &  $N_\text{cfg}$\\
			\hline
			32I             & $32^3 \times 64$ & 0.0828(3)   & 300            &  0.493    & 305\\
			48If            & $48^3 \times 96$ & 0.0711(3)   & 278            &  0.410    & 205\\
		\end{tabular}
	\end{ruledtabular}
\end{table}
\begin{table}[t]
	\caption{\label{tab:mass_table} Meson interpolation operators for different $J^{PC}$ quantum numbers
		used in this work and the masses collected from PDG~\cite{Zyla:2020zbs} for $1S,~1P$ charmonia,
		where $D$ is lattice derivative operator and $B$ is the chromomagnetic gluon field
		from clover discretization of field strength. }
	\begin{ruledtabular}
		\begin{tabular}{cccc}
			meson                & $J^{PC}$ &  operator                                               &  mass(GeV)~\cite{Zyla:2020zbs}
			\\\hline
			$\eta_c({}^1S_0)$    & $0^{-+}$ &  $\gamma_5$                                         & 2.984  \\
			$J/\psi({}^3S_1)$    & $1^{--}$ &  $\gamma_i$                                         & 3.097  \\
			&&&\\
			$\chi_{c0}({}^3P_0)$ & $0^{++}$ &  ${\cal I}$                                         & 3.415  \\
			$\chi_{c1}({}^3P_1)$ & $1^{++}$ &  $\gamma_5\gamma_i$                                 & 3.511  \\
			$h_c({}^1P_0)$       & $1^{+-}$ &  $\gamma_4\gamma_5\gamma_i$                         & 3.525  \\
			$\chi_{c2}({}^3P_2)$ & $2^{++}$ &  $|\epsilon_{ijk}|\gamma_{j}D_k$                    & 3.556  \\
			&&&\\
			$\eta_{c2}({}^1D_2)$ & $2^{-+}$ &  $\epsilon_{ijk}\gamma_4\gamma_5\gamma_{j}D_k$      & ---    \\
			---                  & $1^{-+}$ &  $\epsilon_{ijk}\gamma_{j}B_k$                      & ---    \\
		\end{tabular}
	\end{ruledtabular}
\end{table}
\begin{figure*}[t]
	\centering
	\includegraphics[width=1\linewidth]{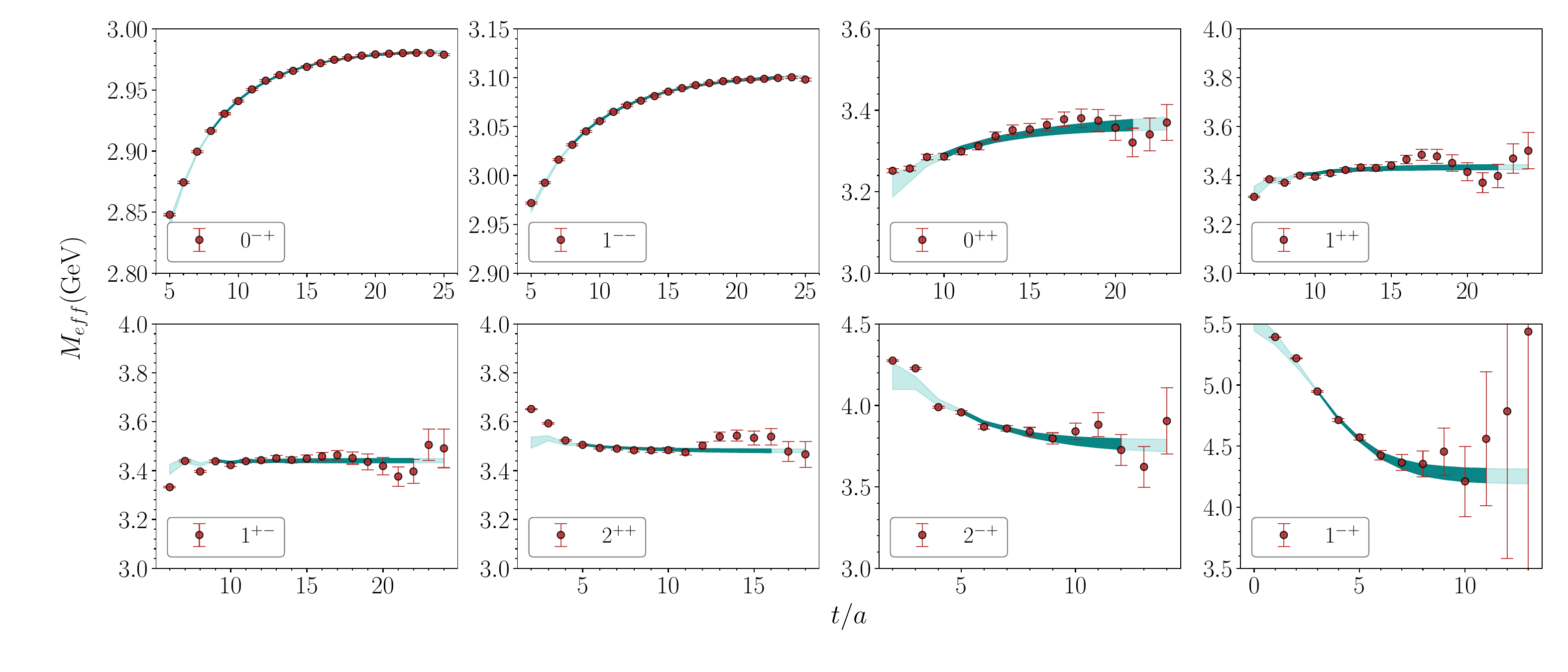}
	\includegraphics[width=1\linewidth]{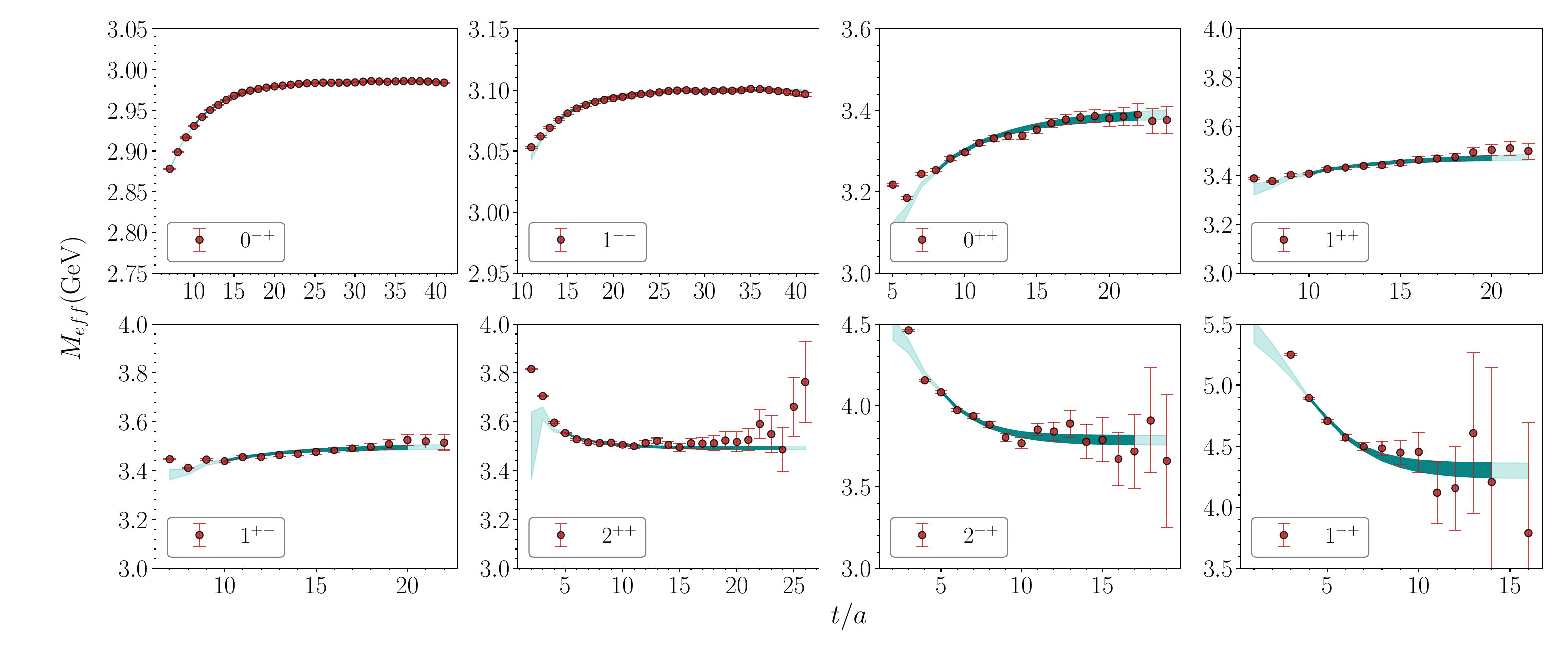}
	\caption{Effective mass as defined in Eq.~(\ref{eq:effem}) from simulation data (points) and from
		$C_2(t)$ of Eq.~(\ref{eq:fit}) with best fit parameters (color band) for various $J^{PC}$ quantum numbers
		on $32^3\times 64$ (top) and $48^3\times 96$ (bottom) configuration.
		The points with error bar are simulated data with jackknife estimated error,
		the light color band shows the fitted result with best fit parameters,
		and the dark color band shows the fitting range.}
	\label{fig:fit_mass}
\end{figure*}
\begin{figure*}[t]
	\centering
	\includegraphics[width=1\linewidth]{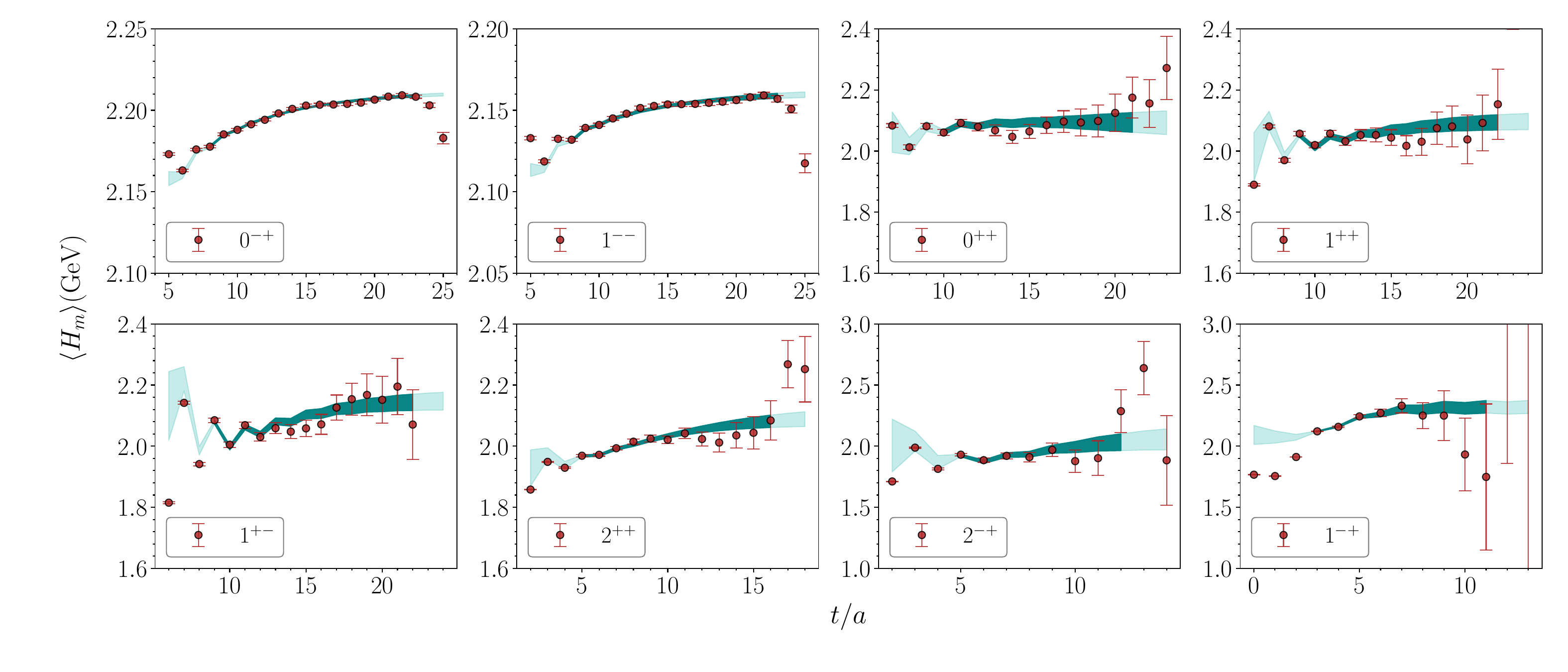}
	\includegraphics[width=1\linewidth]{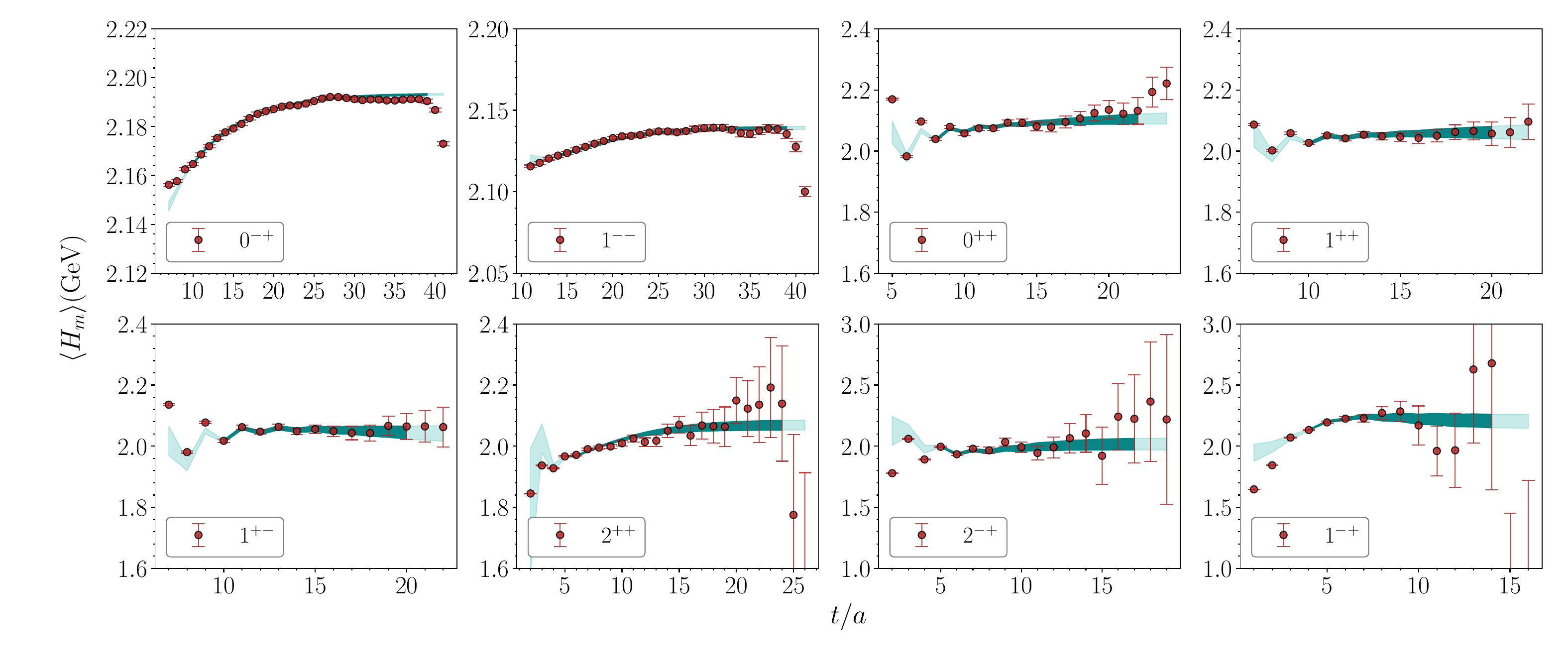}
	\caption{Effective matrix element of valence charm quark mass contribution $\langle H_m \rangle(t)$ as defined in Eq.~(\ref{eq:matrix_element}) from 
	simulation data (points) and from $C_3(t,H_m)$ of Eq.~(\ref{eq:fit}) with best fit parameters (color band) for various $J^{PC}$ quantum numbers
	on $32^3\times 64$ (top) and $48^3\times 96$ (bottom) configuration.
	The points with error bar are simulated data with jackknife estimated error,
	the light color band shows the fitted result with best fit parameters,
	and the dark color band shows the fitting range.}
	\label{fig:fit_hm}
\end{figure*}
\begin{figure*}[t]
	\centering
	\includegraphics[width=1\linewidth]{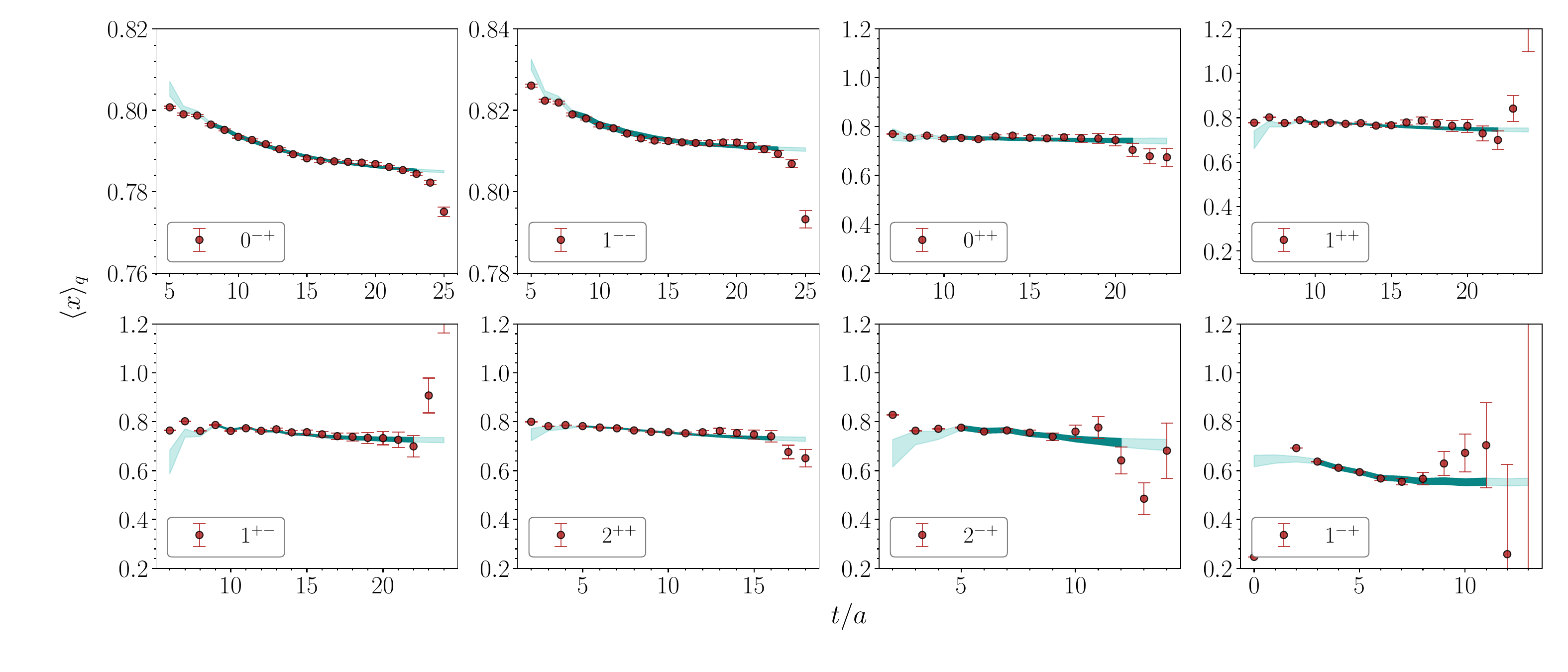}
	\includegraphics[width=1\linewidth]{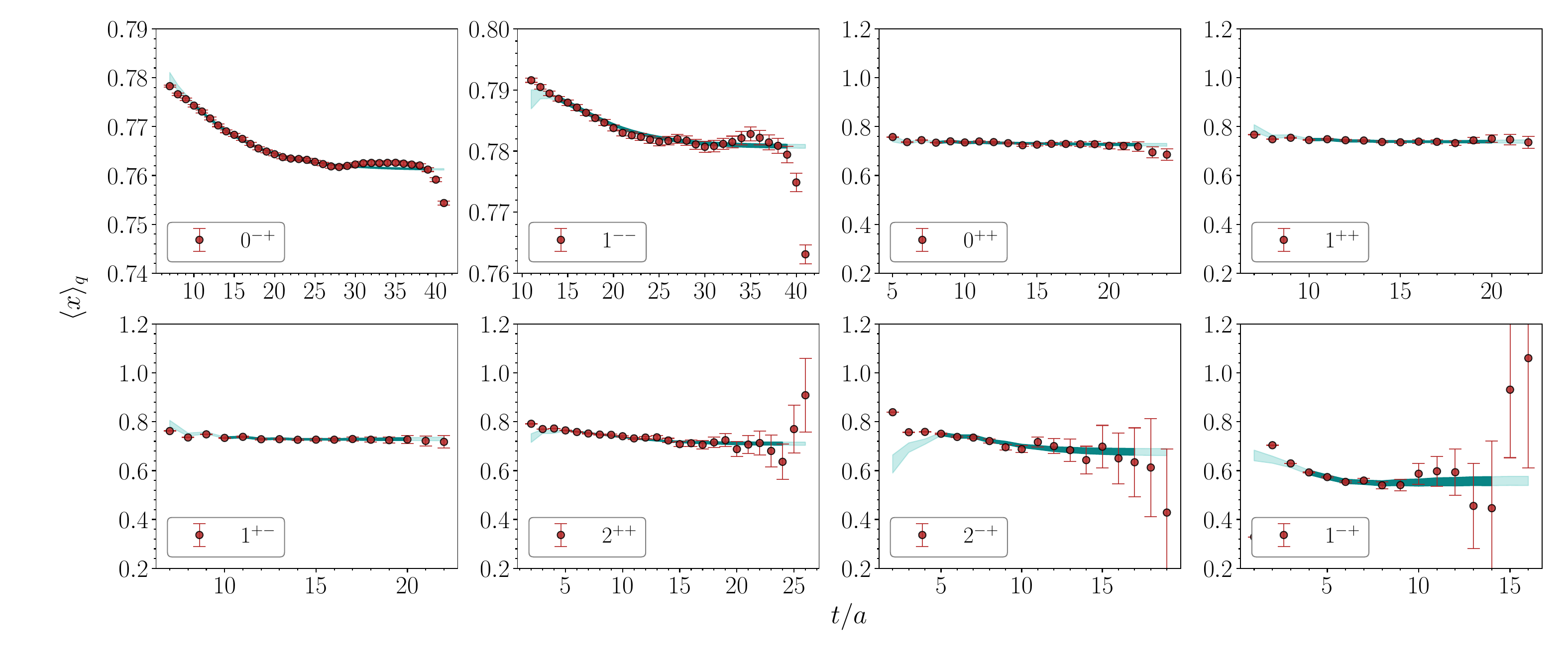}
	\caption{Effective matrix element of valence charm quark momentum fraction $\langle x \rangle_q(t)$ as defined in Eq.~(\ref{eq:matrix_element}) from 
	simulation data (points) and from $C_3(t,x)$ of Eq.~(\ref{eq:fit}) with best fit parameters for various $J^{PC}$ quantum numbers
	on $32^3\times 64$ (top) and $48^3\times 96$ (bottom) configuration.
	The points with error bar are simulated data with jackknife estimated error,
	the light color band shows the fitted result with best fit parameters,
	and the dark color band shows the fitting range.}
	\label{fig:fit_xq}
\end{figure*}
Following the strategy proposed in Ref.~\cite{Yang:2018nqn},
we will calculate the hadron mass $M$, the quark mass contribution $\langle H_m\rangle$,
the quark momentum fraction $\langle x \rangle_q$ and obtain the total quark contribution through the relation,
\begin{align}\label{eq:hq}
\langle H_q^R \rangle =\langle H_E^R \rangle + \langle H_m \rangle = \frac{3}{4} \langle x \rangle^R_q M+\frac{1}{4}\langle H_m\rangle,
\end{align}
where superscript $R$ denote a renormalization scheme at a certain energy scale.
Thus the gluon energy $\langle H_g^R\rangle$ and the QCD trace anomaly part $\langle H_a\rangle$
can be obtained from $M=\langle H_m\rangle+\langle H_a\rangle$ and Eq.~(\ref{eq:total}). In order to extract the masses of charmoniumlike states and study their decompositions,
we calculate the two-point functions $C_2(t)$ involving the operators in Table~\ref{tab:mass_table},
and also the corresponding three-point functions $C_3(t)$ relevant to $\langle H_m\rangle$ and $\langle x \rangle_q$.

For the conventional charmonia, $C_2(t)$ is obtained by the expression
\begin{equation}\label{eq:2pt}
C_2(t,\Gamma)=\langle \gamma_5S^{\dagger}_w( \vec{y},t;0)\gamma_5 \Gamma S_w(\vec{y},t; 0) \Gamma\rangle,
\end{equation}
where $S_w(\vec{y},t; 0)=\sum_{\vec{x}}S(\vec{y},t; \vec{x},0)$ is the Coulomb gauge fixed wall source propagator and $S(\vec{y},t; \vec{x},0)$
is the quark propagator from $(\vec{x},0)$ to $(\vec{y},t)$. Note that the required summation over the spatial indices
is performed implicitly. For the $1^{-+}$ state, one of the $S_w$ terms used above is modified
into $S_w^{B_i}=\sum_{\vec{x}}S(\vec{y},t; \vec{x},0)B_i(\vec{x})$, and the chromomagnetic field strength $B_k$
should be inserted in the sink interpolation field to ensure the correct quantum number. The $2^{\pm+}$ state
can be obtained similarly using the propagator with a derivative in both the source and sink interpolation fields. In order to monitor the $t$-behaviors of $C_2(t,\Gamma)$, we define the effective masses $M_{eff}(t)$ in all the symmetry channels as usual
\begin{equation}\label{eq:effem}
M_{eff}(t)\equiv \ln\frac{C_2(t,\Gamma)}{C_2(t+1,\Gamma)} \ _{\overrightarrow{t\rightarrow \infty}} M,
\end{equation}
and plot them in Fig.~\ref{fig:fit_mass}, where the upper part is for the 32I ensemble and the lower part is for the 48If ensemble (the gauge ensemble parameters can be found in Table~\ref{table:ensemble}). 
It is seen that signal-to-noise ratio is the best for $1S$ states, while the relative errors of the correlation function 
in the $1^{-+}$ channel become very large beyond $t/a=10$. On the other hand, in the small-$t$ range 
of the $M_{eff}(t)$, there appear clear zigzag behaviors, which can be attributed 
to the contribution from the unphysical modes of the domain-wall sea quarks~\cite{Liang:2013eoa}. 
As will be addressed below, this kind of contribution should be taken care of when we extract the masses 
and matrix elements from the correlation function involving the data in the small-$t$ region.

\begin{table*}[t]
	\caption{Fitted ground state hadron mass $M$, quark mass contribution $\langle H_m \rangle$
		and quark momentum fraction $\langle x \rangle_q$ in physical units on two ensembles,
		where the fitting range and $\chi^2$/d.o.f. are also shown in the table.}
	\label{table:fit_result}
	\begin{ruledtabular}
		\begin{tabular}{cllllrl}
			&     $J^{PC}$    & $M$(GeV)         &   $\langle H_m \rangle$(GeV)   &  $\langle x \rangle_q$   & [$t_{min}$ - $t_{max}$]  &  $\chi^2$/d.o.f.  \\
			\hline
			\multirow{8}{*}{$32^3\times 64$}
			&    $0^{-+}$     & 2.983(01)      &   2.212(01)                   &  0.784(01)             &  8 - 23                  &  1.46  \\
			&    $1^{--}$     & 3.104(02)       &   2.162(02)                   &  0.810(01)             &  8 - 23                  &  1.26  \\
			&    $0^{++}$     & 3.375(24)        &   2.087(70)                    &  0.735(22)               & 10 - 21                  &  1.02  \\
			&    $1^{++}$     & 3.434(11)        &   2.101(30)                    &  0.742(11)               &  9 - 22                  &  1.20  \\
			&    $1^{+-}$     & 3.441(09)       &   2.152(33)                    &  0.723(12)               &  9 - 22                  &  1.39  \\
			&    $2^{++}$     & 3.480(08)       &   2.100(35)                    &  0.722(15)               &  5 - 16                  &  1.02  \\
			&    $2^{-+}$     & 3.747(44)        &   2.08(11)                     &  0.694(32)               &  5 - 12                  &  0.84  \\
			&    $1^{-+}$     & 4.250(61)        &   2.317(53)                    &  0.553(15)               &  3 - 11                  &  1.49  \\
			\hline
			&&&&&&\\
			\multirow{8}{*}{$48^3\times 96$}
			&    $0^{-+}$     & 2.985(01)      &   2.193(01)                  &  0.761(01)             & 10 - 39                  &  1.39  \\
			&    $1^{--}$     & 3.100(01)      &   2.139(01)                  &  0.781(01)             & 14 - 39                  &  1.44  \\
			&    $0^{++}$     & 3.396(17)        &   2.113(25)                    &  0.722(08)              &  8 - 22                  &  1.50  \\
			&    $1^{++}$     & 3.480(18)        &   2.063(38)                    &  0.741(14)               & 10 - 20                  &  1.35  \\
			&    $1^{+-}$     & 3.500(14)        &   2.030(44)                    &  0.732(13)               & 10 - 20                  &  0.97  \\
			&    $2^{++}$     & 3.492(07)       &   2.070(17)                    &  0.710(07)              &  5 - 24                  &  1.55  \\
			&    $2^{-+}$     & 3.788(30)        &   2.019(49)                    &  0.675(15)               &  5 - 17                  &  1.33  \\
			&    $1^{-+}$     & 4.296(64)        &   2.202(59)                    &  0.559(19)               &  4 - 14                  &  0.96  \\
		\end{tabular}
	\end{ruledtabular}
	
\end{table*}

For the three-point functions, we use the summed current sequential source method~\cite{Chang:2018uxx} to suppress excited state contamination. To be specific, the current-summed three-point function is expressed as
\begin{align}\label{eq:3pt}
    C_3(t,\Gamma,O)=\langle \gamma_5 S^{\dagger}_w(\vec{y},t;0)\gamma_5 \Gamma \tilde{S}_{c}(\vec{y},t;0;O) \Gamma\rangle,
\end{align}
where the current sequential propagator $\tilde{S}_c(\vec{y},t;0;O)$ is defined as
\begin{align}
    \tilde{S}_{c}(\vec{y},t;0;H_m)=&m_c\sum_{\vec{x},t'}S(\vec{y},t;\vec{x},t')S_w(\vec{x},t';0),\\\nonumber
    \tilde{S}_{c}(\vec{y},t; 0;x)=&\sum_{\vec{x},t'}S(\vec{y},t;\vec{x},t')(D_4\gamma_4-\frac{1}{3}\sum_i D_i\gamma_i)\nonumber\\
    &\quad S_w(\vec{x},t;0),
\end{align}
for $O=H_m$ and $O=x$, respectively, $m_c$ is the bare charm quark mass, and the current insertion time $t'$ is summed over all time slice (0 to $T-1$), 
which results in the exponential suppression of excited state contribution and benefits the fitting by the ability to explore all the source-sink separation of three-point functions.
In the $1^{-+}$ channel, we need to replace one of the $S_w$ terms in Eq.~(\ref{eq:3pt})
by the $S_w^{B_i}$ (Note that $S_w^{B_i}$ should be symmetric with respect to the interchange of spatial coordinates to ensure the correct $P$-parity). Similar to the effective masses, we define the effective matrix elements as 
\begin{align}\label{eq:matrix_element}
&\langle H_m\rangle(t)= R(t,H_m)-R(t-1,H_m) ,\nonumber\\
&M\langle x\rangle_q(t)= R(t,x)-R(t-1,x).
\end{align}
where $R(t,O)$ is the ratio function $R(t,O)\equiv C_3(t,\Gamma,O)/C_2(t,\Gamma)$. 
It can be easily verified that $\langle H_m\rangle(t)$ and $M\langle x\rangle_q(t)$ is dominated 
by the related matrix elements of the ground state and independent of $t$ in the large $t$ limit. 
The effective matrix elements $\langle H_m \rangle(t)$ and $\langle x \rangle_q(t)$ in all the $J^{PC}$ channels 
are plotted in Fig.~\ref{fig:fit_hm},~\ref{fig:fit_xq}, respectively. In each figure, 
the upper part is for the 32I ensemble and the lower part is for the 48If ensemble. 
Obviously, there are no plateaus showing up in the time windows for most of the effective matrix elements. 
This implies that there are substantial contaminations from higher states. In the mean time, 
the zigzag behaviors are more apparent in the small $t$ region and are essentially inherited 
from the two-point functions $C_2(t,\Gamma)$ (it is understood that the unphysical modes 
of domain wall sea quarks contribute little to the three point functions involved in this work). 

Since the spectrum is common for $C_2(t,\Gamma)$ and $C_3(t,\Gamma,O)$ in each $J^{PC}$ channel, we parametrized them in the following function forms 
\begin{align}\label{eq:fit}
&C_3(t,H_m)=e^{-Mt}\left(A_0 t \langle H_m\rangle + A_2e^{-\delta m t} + A_3t e^{-\delta m t}+A_4\right),\nonumber\\
&C_3(t,x)=e^{-Mt}\left(A_0t M \langle x\rangle_q +A_5e^{-\delta m t} + A_6t e^{-\delta m t}+A_7\right),\nonumber\\
&C_2(t)=A_0e^{-M t} (1+A_1e^{-\delta m t}) + W(-1)^{t}e^{-\tilde{M}t},
\end{align}
where $M$ is the mass of the ground state, the $e^{-\delta m t}$ terms are introduced to account 
for the contamination from higher states, $\delta m$, $A_{0,...,7}$ are free parameters 
and the $(-1)^{t}$ term is added to account for the oscillatory behavior related to domain wall fermions. 
These function forms facilitate us to use the data of the correlation functions at earlier time slices 
to suppress the statistical uncertainties. In each $J^{PC}$ channel, we carry out a joint correlated fit 
to $C_3(t,H_m)$, $C_3(t,x)$ and $C_2(t)$ using these function forms, 
thus the quantities $M$, $\langle H_m\rangle$ and $\langle x \rangle_q$ can be extracted simultaneously. 
This fitting strategy works well practically, as illustrated by the shaded bands in each plot 
in Fig.~\ref{fig:fit_mass},~\ref{fig:fit_hm},~\ref{fig:fit_xq}, where the dark color band is plotted using 
the fitted parameters in the specific time range and the light color band shows the extrapolation result. 
Obviously, the function forms in Eq.~(\ref{eq:fit}) describe the data very well. The fitted results 
of $M$, $\langle H_m\rangle$ and $\langle x \rangle_q$
are tabulated in Table~\ref{table:fit_result} along with the fitting window $[t_\mathrm{min},t_\mathrm{max}]$ 
and the $\chi^2$ per degree of freedom ($\chi^2/\mathrm{d.o.f.}$) for each fit, 
where the masses $M$ and $\langle H_m\rangle$ are converted into the values in the physical units. 

In principle, the renormalization of $\langle x\rangle_q$ should be considered. First, $\langle x\rangle_q$ 
can mix with the momentum fraction of gluons after the renormalization. However, it is found that the mixing 
from gluon momentum fraction to that of quark is at 1\% level per flavor in a previous work of proton mass decomposition 
at several lattice spacings~\cite{Yang:2018nqn}, thus this mixing effect can be tentative ignored. 
Secondly, the quark momentum fraction itself
is almost multiplicatively renormalizable. In Ref.~\cite{Yang:2018nqn},  this renormalization constant 
is determined to be  $Z_{QQ}(\text{32I})=0.99(5)$ at $a$=0.0828(3) fm, which implies the renormalization effect 
can be very small. We have not obtained $Z_{QQ}$ on 48If yet,  but considering that $\langle x \rangle_q$ 
on 48If differs from that on 32I only by $\sim$4\%, we just present the bare $\langle x \rangle_q$ 
and ignore the renormalization effects in current study.

\section{Discussion}\label{result}
\begin{figure}[t]
	\centering
	\includegraphics[width=1\linewidth]{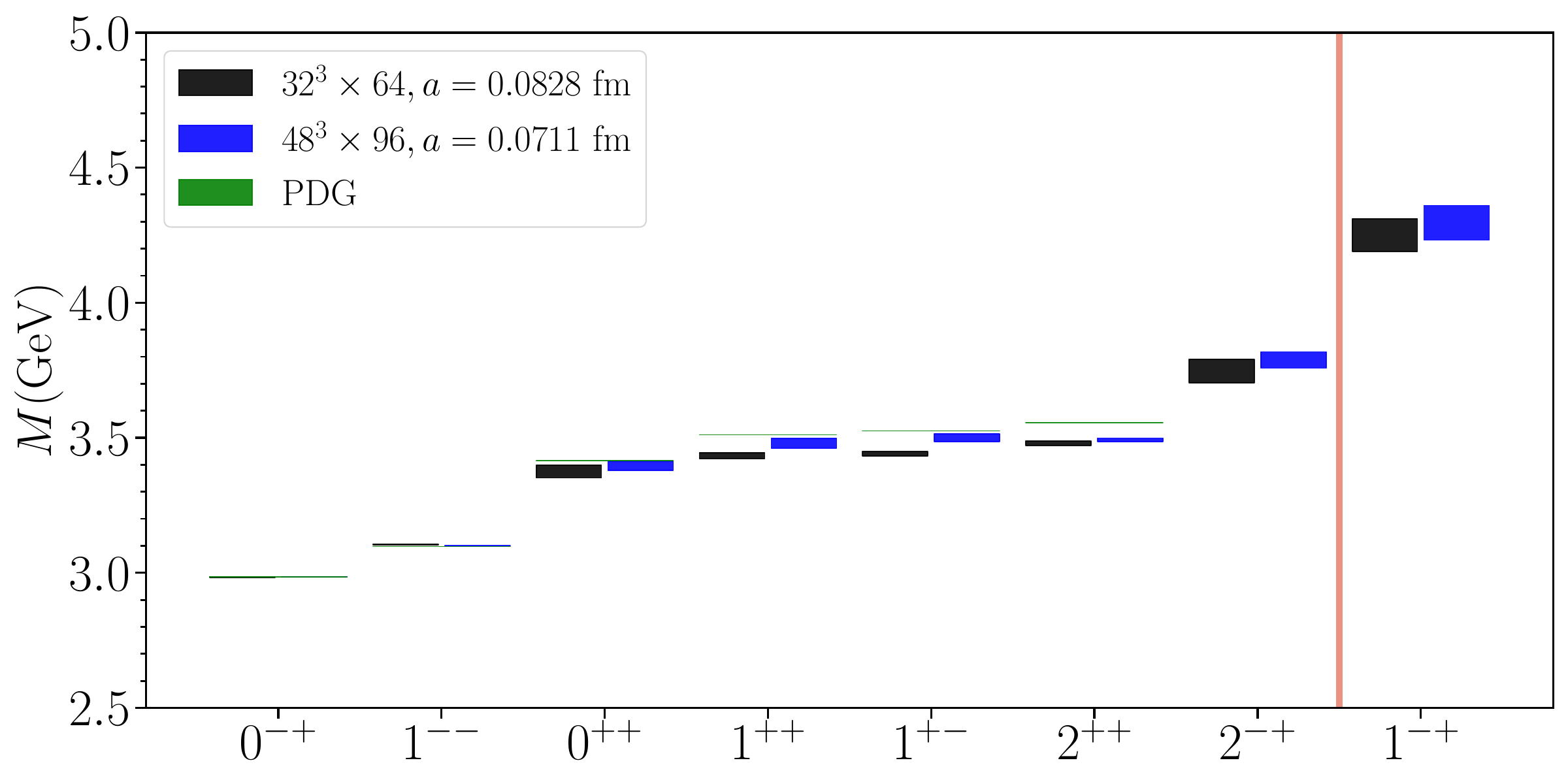}
	\caption{Ground state mass $M$ of various $J^{PC}$ quantum numbers
		on $32^3\times 64$ and $48^3\times 96$ configurations,
		the height of color boxes indicates the statistical errors and
		the experimental value for $1S$ and $1P$ states from PDG are shown for comparison.}
	\label{fig:mass}
\end{figure}

We start with the discussion on the calculated masses of the charmonia and the charmoniumlike state in this work. Our results are plotted in Fig.~\ref{fig:mass} with the black (32I ensemble) and blue boxes (42If ensemble) with the heights of the boxes showing the statistical errors. The experimental values from PDG~\cite{Zyla:2020zbs} are also given in green lines for comparison. By setting the bare charm quark masses with the PDG value of the $J/\psi(1^{--})$ mass, our prediction of the hyperfine-splitting of $J/\psi-\eta_c$ on the two ensembles agrees with the PDG value
within 5\% difference. The masses of the $\{1^{+-}, (0,1,2)^{++}\}$ states are a little lower than those of experimental $1P$ charmonia $\{h_c, \chi_{c0,1,2}\}$ and have mild finite lattice spacing dependences. This deviation may be attributed to the discretization uncertainties that are not tackled in this work.
The ground state mass in $2^{-+}$ channel is determined to be 3.747(44) GeV (32I) and 3.788(30) GeV (48If). Even though the $2^{-+}$ charmonium, which is usually named $\eta_{c2}$ and assigned to be $1^1D_2$ state,
has not been observed in experiments, its mass should be in this range according to the masses of its spin-triplet partners $\psi(3770) (1^3D_1)$, $\psi_2(3823)(1^3D_2)$~\cite{Bhardwaj:2013rmw,Ablikim:2015dlj}
and $\psi_3(3842)(1^3D_3)$~\cite{Aaij:2019evc}. The $1^{-+}$ state, with an exotic quantum number, is around 4.2 to 4.3 GeV, agrees with the result of the latest lattice QCD calculation with dynamical quarks~\cite{Liu:2012ze}.

\begin{figure}[t!]
    \centering
    \includegraphics[width=1\linewidth]{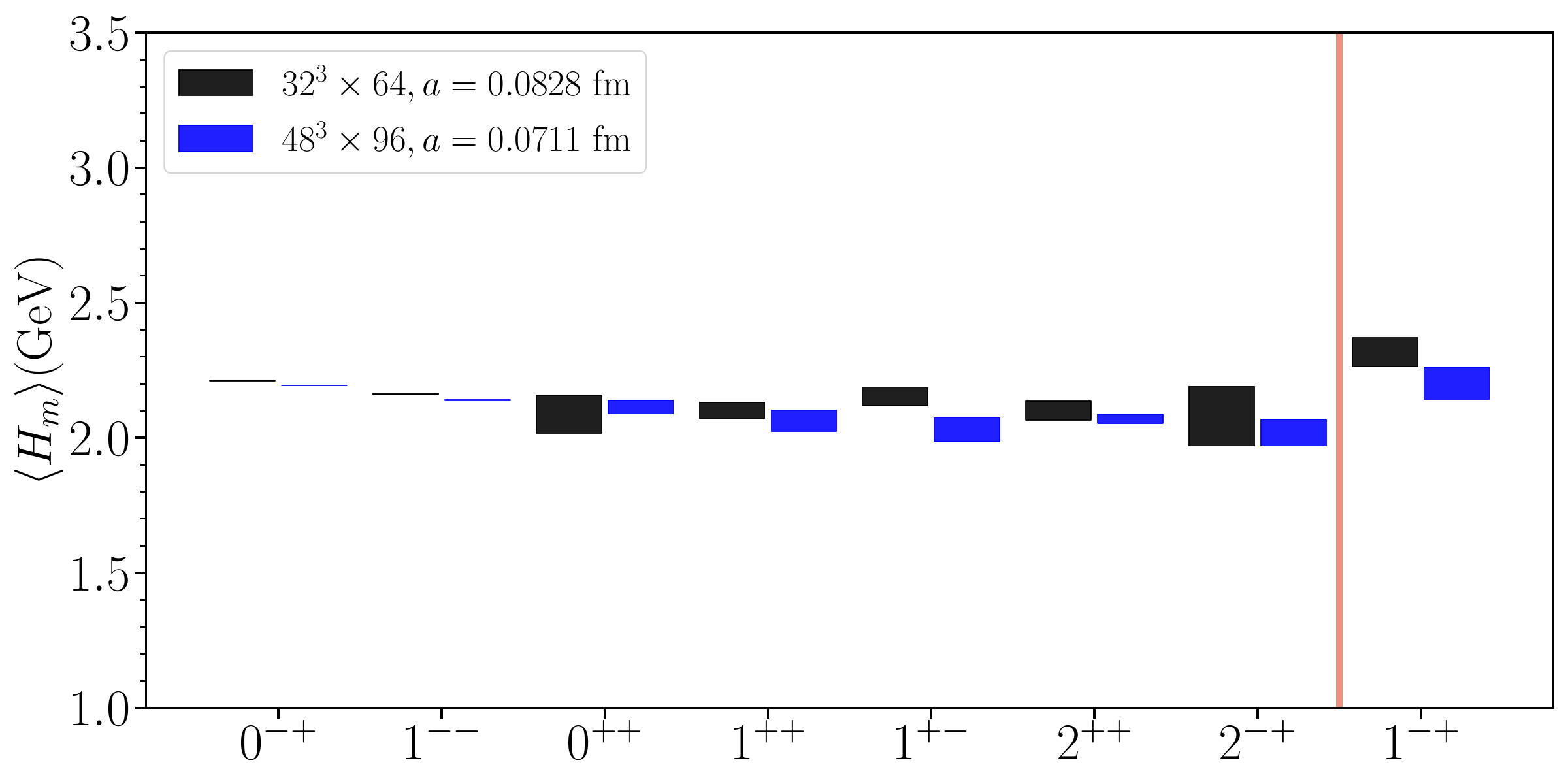}
    \includegraphics[width=1\linewidth]{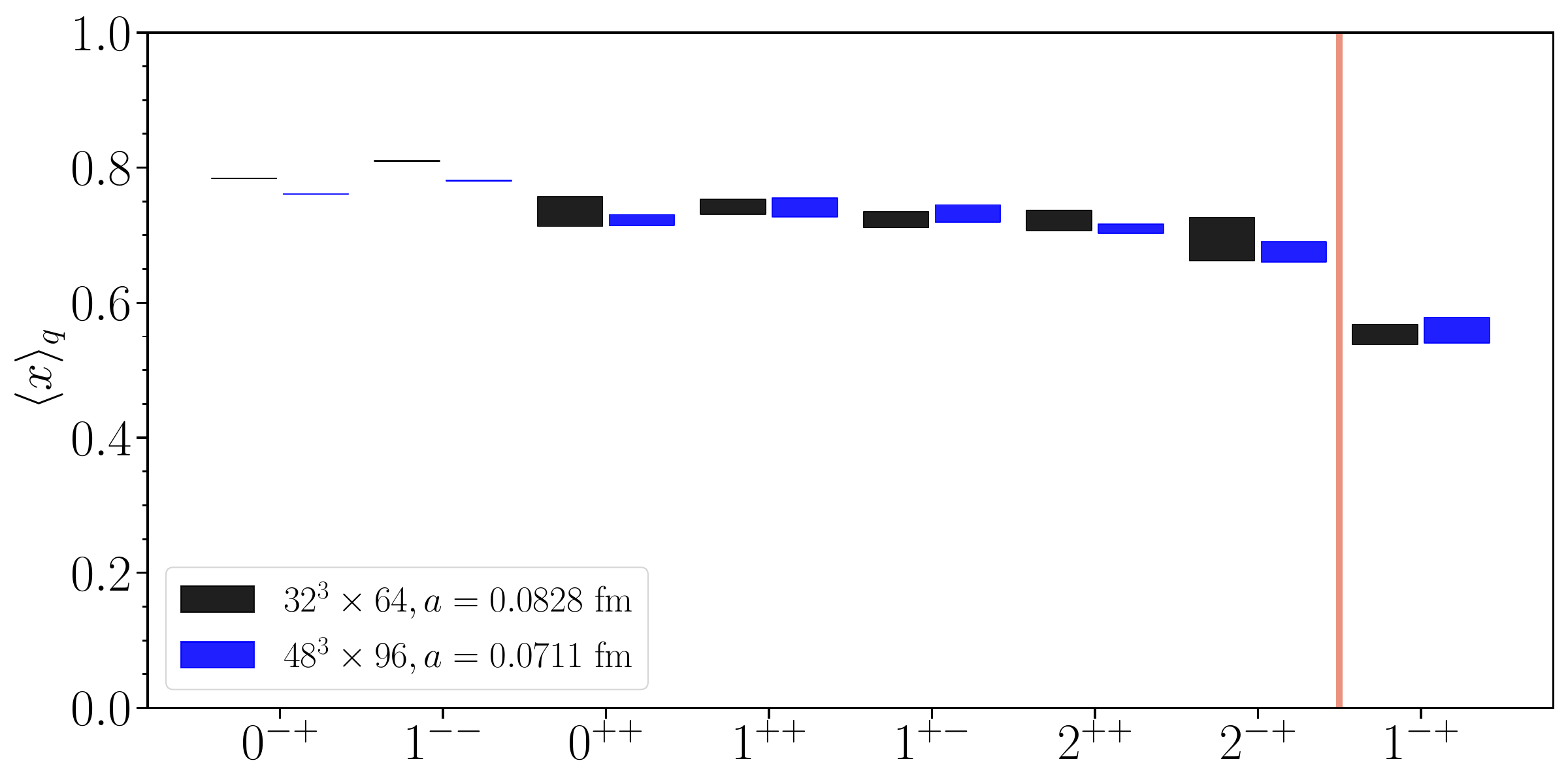}
    \caption{Ground state $\langle H_m \rangle$ and $\langle x \rangle_q$ of valence charm quark of various $J^{PC}$ quantum numbers
    on $32^3\times 64$ and $48^3\times 96$ configuration.}
    \label{fig:x}
\end{figure}

\begin{figure}[t]
    \centering
    \includegraphics[width=1\linewidth]{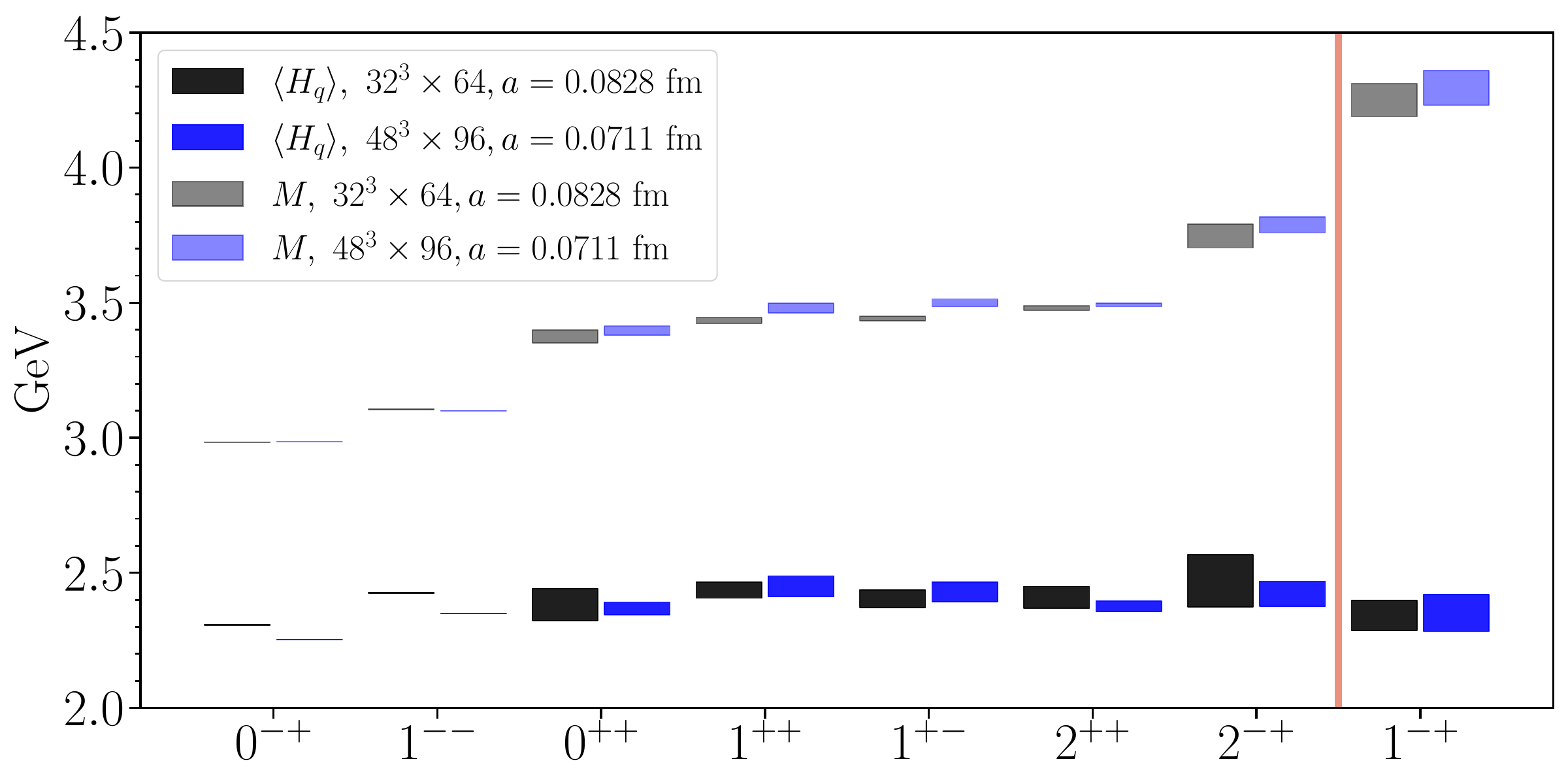}
    \caption{Combined valence charm quark contribution $\langle H_q \rangle$ in ground state of various $J^{PC}$ quantum numbers
    on $32^3\times 64$ and $48^3\times 96$ configuration, the extracted mass $M$ are also shown for comparison.}
    \label{fig:H_q}
\end{figure}

Quark mass contribution $\langle H_m\rangle $ in different states are illustrated in the upper panel of Fig.~\ref{fig:x}.
For all the states, $\langle H_m \rangle$ is around 2.0 to 2.2 GeV.
Note that $\langle H_m\rangle$ here only includes the contribution from valence charm quarks.
The contribution from charm sea quarks is subtle since the 2+1 flavor ensemble is used
in this work and we do not have any charm sea in the configurations,
but its contribution $\langle H_m^{c,sea}\rangle$ can be estimated from the charm valence quark contribution
$\langle H_m^{c,v} \rangle$ (the actual $\langle H_m \rangle$ we obtained)
based on the heavy quark expansion~\cite{Shifman:1978zn},
\begin{align}
\langle H_m^{c,sea} \rangle =\frac{2}{27}\big(\frac{1}{1+\gamma_m(\mu)}M-\langle H_m^{c,v} \rangle\big)+{\cal O}(\alpha_s),
\end{align}
where $\gamma_m(\mu)\approx 2\alpha(\mu)/\pi$ is anomalous dimension of the quark mass,
and if we take $\alpha(\mu=m_c)\approx 0.37$~\cite{Maezawa:2016vgv}, then $\langle H_m^{c,sea} \rangle$ is less than
100 MeV for all these states, therefore we just ignore the charm sea quark contribution in this study.

The light and strange sea quark contributions to $\langle H_m\rangle$ can be estimated through the Feynman-Hellman theorem. Using the light quark mass dependence of charmonium masses obtained in Ref.~\cite{Yang:2014sea},
 the matrix element $\langle \int d^3x \bar{\psi}(x)\psi(x)\rangle$
is approximately 0.3 at $\overline{\textrm{MS}}$ 2 GeV for $u,d$ quarks. This quantity for the strange quark should not be larger, and then
 $u,d,s$ sea quark contribution to $\langle H_m\rangle$ would be not larger than 40 MeV in total and can be ignored temporarily.

On the other hand, estimation of the sea quark momentum fraction is relatively nontrivial.
We know that the light sea quark momentum fraction is a few percent in the nucleon and that of the heavy quark is further suppressed by the quark mass
and then would be negligible~\cite{Alexandrou:2020sml}.

According to the relation $M=\langle H_m\rangle +\langle H_a\rangle$, one can conclude that the mass
differences among $1S$, $1P$, $1D$ charmonia and $1^{-+}$ charmoniumlike state mainly
from the contribution of the QCD trace anomaly part $\langle H_a\rangle$.
The quark relevant part of $\langle H_a\rangle$ is $\gamma_m \langle H_m\rangle$ as shown in Eq.~(\ref{eq:decomp}).
Since $\gamma_m(\mu)$ is common for all states and $\langle H_m\rangle$ is scale independent,
we can further claim that the mass differences are dominated
by the gluon components of charmoniumlike states.

It is interesting to see that, as shown in the lower panel of Fig.~\ref{fig:x},
the momentum fraction $\langle x\rangle_q$ of the valence charm quark and antiquark is
in the range of 70\%-80\% and does not change much for conventional $1S$, $1P$ and $1D$ charmonia,
while it is only approximately 55\% for the $1^{-+}$ state. This is a striking difference,
which means there is a roughly 20\% more momentum fraction carried by other degrees of freedom,
such as gluons and $u,d,s$ sea quarks in the $1^{-+}$ state, in comparison with that of $J/\psi$
and $\eta_c$. This surplus amount of momentum fraction accounts for approximately 800 MeV
of the mass of the $1^{-+}$ state, given its mass around 4.3 GeV.
If the $1^{-+}$ state is interpreted phenomenologically as a $c\bar{c}g$ hybrid,
then this portion of mass of 800 MeV is close to the mass of a constituent gluon
in the constituent model for glueballs and hybrids. However, one should be cautious to make this argument
since a $c\bar{c}g$ hybrid is ill-defined in QCD due to gluon-$q\bar{q}$ transition within a hadron system,
which also allows possible $c\bar{c}q\bar{q}\ldots $ configurations. Whatever this $1^{-+}$ charmoniumlike state is,
the striking difference of its $\langle x\rangle_q$ from the conventional charmonia may signals its special inner structure.

\section{Summary}\label{summary}
We investigated mass decomposition of conventional $1S,1P,1D$ charmonia and
exotic $1^{-+}$ charmoniumlike state. It is found that quark mass contribution $\langle H_m\rangle$ of valence charm quark is around 2.0 to 2.2 GeV for these states,
and the valence charm quark momentum fraction $\langle x\rangle_q$ of the $1^{-+}$ state is $\sim$0.55 while that in the other conventional charmonia is around 0.7 to 0.8.
Moreover, according to Eq.~(\ref{eq:hq}), the combined quark contribution $\langle H_q\rangle$ of the valence charm quark
can be expressed in terms of $\langle x\rangle_q$ and $\langle H_m\rangle$, whose values for different states
are shown in Fig.~\ref{fig:H_q} along with the masses of charmoniumlike states (light color boxes). It is clearly seen that the $\langle H_q \rangle$ in all the states
are close to each other and around 2.3-2.5 GeV. The mass sum rule in Eq.~(\ref{eq:total}) implies again
that the gluon contribution to the masses is $M-\langle H_q\rangle$
(except for a small portion $\frac{1}{4}\gamma_m(\mu)\langle H_m\rangle)$ that gives the major differences
among charmoniumlike states. Thus our calculation provides an evidence of the significant gluon contribution in the charmonium states.

In summary, we perform an exploratory study on mass decomposition of charmoniumlike states
and demonstrate the possibility to understand the charmonium and the other meson masses through the structure calculation,
besides the standard spectrum analysis. Similar calculation can be applied to the other mesons with heavier or lighter quark mass,
to uncover the phenomenology meaning of different mass components. The unstable particles like the $\rho$, $\sigma$ meson, XYZ particles,
or even light nucleus also worth a similar investigation with different volumes,
and such an investigation would shed light on properties of these particles accompanying the state-of-arts scattering studies.

In this work, we focus on the structure property of charmoniumlike states, and the finite-volume effect is ignored. As the first step of such study, we believe that this effect would not change our current conclusion.
However for a precise theoretical prediction, it should be considered in detail and we will investigate it in our future works.

\begin{acknowledgements}
We thank the RBC and UKQCD collaborations for providing us their DWF gauge configurations.
The calculations were performed using the GWU-code~\cite{Alexandru:2011ee,Alexandru:2011sc} through HIP programming model~\cite{Bi:2020wpt}.
This work is supported by the National Key Research and Development Program of China (No.2017YFB0203202),
the Strategic Priority Research Program of Chinese Academy of Sciences (No.XDC01040100 and No.XDB34030300),
and the Natural Science Foundation of China under grant No.11935017, No.11975127 and No.12070131001 (CRC 110 by DFG and NNSFC).
The computing resources of the Southern China Nuclear Computer center (SCNC) and HPC Cluster of ITP-CAS are acknowledged.
Y. Chen is also supported by the CAS Center for Excellence in Particle Physics (CCEPP).
P. Sun is also supported by Jiangsu Specially Appointed Professor Program.
\end{acknowledgements}

\bibliography{reference}

\begin{thebibliography}{25}%
\makeatletter
\providecommand \@ifxundefined [1]{%
 \@ifx{#1\undefined}
}%
\providecommand \@ifnum [1]{%
 \ifnum #1\expandafter \@firstoftwo
 \else \expandafter \@secondoftwo
 \fi
}%
\providecommand \@ifx [1]{%
 \ifx #1\expandafter \@firstoftwo
 \else \expandafter \@secondoftwo
 \fi
}%
\providecommand \natexlab [1]{#1}%
\providecommand \enquote  [1]{``#1''}%
\providecommand \bibnamefont  [1]{#1}%
\providecommand \bibfnamefont [1]{#1}%
\providecommand \citenamefont [1]{#1}%
\providecommand \href@noop [0]{\@secondoftwo}%
\providecommand \href [0]{\begingroup \@sanitize@url \@href}%
\providecommand \@href[1]{\@@startlink{#1}\@@href}%
\providecommand \@@href[1]{\endgroup#1\@@endlink}%
\providecommand \@sanitize@url [0]{\catcode `\\12\catcode `\$12\catcode
  `\&12\catcode `\#12\catcode `\^12\catcode `\_12\catcode `\%12\relax}%
\providecommand \@@startlink[1]{}%
\providecommand \@@endlink[0]{}%
\providecommand \url  [0]{\begingroup\@sanitize@url \@url }%
\providecommand \@url [1]{\endgroup\@href {#1}{\urlprefix }}%
\providecommand \urlprefix  [0]{URL }%
\providecommand \Eprint [0]{\href }%
\providecommand \doibase [0]{http://dx.doi.org/}%
\providecommand \selectlanguage [0]{\@gobble}%
\providecommand \bibinfo  [0]{\@secondoftwo}%
\providecommand \bibfield  [0]{\@secondoftwo}%
\providecommand \translation [1]{[#1]}%
\providecommand \BibitemOpen [0]{}%
\providecommand \bibitemStop [0]{}%
\providecommand \bibitemNoStop [0]{.\EOS\space}%
\providecommand \EOS [0]{\spacefactor3000\relax}%
\providecommand \BibitemShut  [1]{\csname bibitem#1\endcsname}%
\let\auto@bib@innerbib\@empty
\bibitem [{\citenamefont {Ji}(1995)}]{Ji:1994av}%
  \BibitemOpen
  \bibfield  {author} {\bibinfo {author} {\bibfnamefont {X.-D.}\ \bibnamefont
  {Ji}},\ }\href {\doibase 10.1103/PhysRevLett.74.1071} {\bibfield  {journal}
  {\bibinfo  {journal} {Phys. Rev. Lett.}\ }\textbf {\bibinfo {volume} {74}},\
  \bibinfo {pages} {1071} (\bibinfo {year} {1995})},\ \Eprint
  {http://arxiv.org/abs/hep-ph/9410274} {arXiv:hep-ph/9410274 [hep-ph]}
  \BibitemShut {NoStop}%
\bibitem [{\citenamefont {Shifman}\ \emph {et~al.}(1978)\citenamefont
  {Shifman}, \citenamefont {Vainshtein},\ and\ \citenamefont
  {Zakharov}}]{Shifman:1978zn}%
  \BibitemOpen
  \bibfield  {author} {\bibinfo {author} {\bibfnamefont {M.~A.}\ \bibnamefont
  {Shifman}}, \bibinfo {author} {\bibfnamefont {A.~I.}\ \bibnamefont
  {Vainshtein}}, \ and\ \bibinfo {author} {\bibfnamefont {V.~I.}\ \bibnamefont
  {Zakharov}},\ }\href {\doibase 10.1016/0370-2693(78)90481-1} {\bibfield
  {journal} {\bibinfo  {journal} {Phys. Lett.}\ }\textbf {\bibinfo {volume}
  {B78}},\ \bibinfo {pages} {443} (\bibinfo {year} {1978})}\BibitemShut
  {NoStop}%
\bibitem [{\citenamefont {Lorc{\'e}}(2018)}]{Lorce:2017xzd}%
  \BibitemOpen
  \bibfield  {author} {\bibinfo {author} {\bibfnamefont {C.}~\bibnamefont
  {Lorc{\'e}}},\ }\href {\doibase 10.1140/epjc/s10052-018-5561-2} {\bibfield
  {journal} {\bibinfo  {journal} {Eur. Phys. J.}\ }\textbf {\bibinfo {volume}
  {C78}},\ \bibinfo {pages} {120} (\bibinfo {year} {2018})},\ \Eprint
  {http://arxiv.org/abs/1706.05853} {arXiv:1706.05853 [hep-ph]} \BibitemShut
  {NoStop}%
\bibitem [{\citenamefont {Yang}\ \emph {et~al.}(2018)\citenamefont {Yang},
  \citenamefont {Liang}, \citenamefont {Bi}, \citenamefont {Chen},
  \citenamefont {Draper}, \citenamefont {Liu},\ and\ \citenamefont
  {Liu}}]{Yang:2018nqn}%
  \BibitemOpen
  \bibfield  {author} {\bibinfo {author} {\bibfnamefont {Y.-B.}\ \bibnamefont
  {Yang}}, \bibinfo {author} {\bibfnamefont {J.}~\bibnamefont {Liang}},
  \bibinfo {author} {\bibfnamefont {Y.-J.}\ \bibnamefont {Bi}}, \bibinfo
  {author} {\bibfnamefont {Y.}~\bibnamefont {Chen}}, \bibinfo {author}
  {\bibfnamefont {T.}~\bibnamefont {Draper}}, \bibinfo {author} {\bibfnamefont
  {K.-F.}\ \bibnamefont {Liu}}, \ and\ \bibinfo {author} {\bibfnamefont
  {Z.}~\bibnamefont {Liu}},\ }\href {\doibase 10.1103/PhysRevLett.121.212001}
  {\bibfield  {journal} {\bibinfo  {journal} {Phys. Rev. Lett.}\ }\textbf
  {\bibinfo {volume} {121}},\ \bibinfo {pages} {212001} (\bibinfo {year}
  {2018})},\ \Eprint {http://arxiv.org/abs/1808.08677} {arXiv:1808.08677
  [hep-lat]} \BibitemShut {NoStop}%
\bibitem [{\citenamefont {Yang}\ \emph
  {et~al.}(2015{\natexlab{a}})\citenamefont {Yang}, \citenamefont {Chen},
  \citenamefont {Draper}, \citenamefont {Gong}, \citenamefont {Liu},
  \citenamefont {Liu},\ and\ \citenamefont {Ma}}]{Yang:2014xsa}%
  \BibitemOpen
  \bibfield  {author} {\bibinfo {author} {\bibfnamefont {Y.-B.}\ \bibnamefont
  {Yang}}, \bibinfo {author} {\bibfnamefont {Y.}~\bibnamefont {Chen}}, \bibinfo
  {author} {\bibfnamefont {T.}~\bibnamefont {Draper}}, \bibinfo {author}
  {\bibfnamefont {M.}~\bibnamefont {Gong}}, \bibinfo {author} {\bibfnamefont
  {K.-F.}\ \bibnamefont {Liu}}, \bibinfo {author} {\bibfnamefont
  {Z.}~\bibnamefont {Liu}}, \ and\ \bibinfo {author} {\bibfnamefont {J.-P.}\
  \bibnamefont {Ma}},\ }\href {\doibase 10.1103/PhysRevD.91.074516} {\bibfield
  {journal} {\bibinfo  {journal} {Phys. Rev.}\ }\textbf {\bibinfo {volume}
  {D91}},\ \bibinfo {pages} {074516} (\bibinfo {year} {2015}{\natexlab{a}})},\
  \Eprint {http://arxiv.org/abs/1405.4440} {arXiv:1405.4440 [hep-ph]}
  \BibitemShut {NoStop}%
\bibitem [{\citenamefont {Liu}\ \emph {et~al.}(2012)\citenamefont {Liu},
  \citenamefont {Moir}, \citenamefont {Peardon}, \citenamefont {Ryan},
  \citenamefont {Thomas}, \citenamefont {Vilaseca}, \citenamefont {Dudek},
  \citenamefont {Edwards}, \citenamefont {Joo},\ and\ \citenamefont
  {Richards}}]{Liu:2012ze}%
  \BibitemOpen
  \bibfield  {author} {\bibinfo {author} {\bibfnamefont {L.}~\bibnamefont
  {Liu}}, \bibinfo {author} {\bibfnamefont {G.}~\bibnamefont {Moir}}, \bibinfo
  {author} {\bibfnamefont {M.}~\bibnamefont {Peardon}}, \bibinfo {author}
  {\bibfnamefont {S.~M.}\ \bibnamefont {Ryan}}, \bibinfo {author}
  {\bibfnamefont {C.~E.}\ \bibnamefont {Thomas}}, \bibinfo {author}
  {\bibfnamefont {P.}~\bibnamefont {Vilaseca}}, \bibinfo {author}
  {\bibfnamefont {J.~J.}\ \bibnamefont {Dudek}}, \bibinfo {author}
  {\bibfnamefont {R.~G.}\ \bibnamefont {Edwards}}, \bibinfo {author}
  {\bibfnamefont {B.}~\bibnamefont {Joo}}, \ and\ \bibinfo {author}
  {\bibfnamefont {D.~G.}\ \bibnamefont {Richards}} (\bibinfo {collaboration}
  {Hadron Spectrum}),\ }\href {\doibase 10.1007/JHEP07(2012)126} {\bibfield
  {journal} {\bibinfo  {journal} {JHEP}\ }\textbf {\bibinfo {volume} {07}},\
  \bibinfo {pages} {126} (\bibinfo {year} {2012})},\ \Eprint
  {http://arxiv.org/abs/1204.5425} {arXiv:1204.5425 [hep-ph]} \BibitemShut
  {NoStop}%
\bibitem [{\citenamefont {Yang}\ \emph {et~al.}(2013)\citenamefont {Yang},
  \citenamefont {Chen}, \citenamefont {Gui}, \citenamefont {Liu}, \citenamefont
  {Liu}, \citenamefont {Liu}, \citenamefont {Ma},\ and\ \citenamefont
  {Zhang}}]{Yang:2012mya}%
  \BibitemOpen
  \bibfield  {author} {\bibinfo {author} {\bibfnamefont {Y.-B.}\ \bibnamefont
  {Yang}}, \bibinfo {author} {\bibfnamefont {Y.}~\bibnamefont {Chen}}, \bibinfo
  {author} {\bibfnamefont {L.-C.}\ \bibnamefont {Gui}}, \bibinfo {author}
  {\bibfnamefont {C.}~\bibnamefont {Liu}}, \bibinfo {author} {\bibfnamefont
  {Y.-B.}\ \bibnamefont {Liu}}, \bibinfo {author} {\bibfnamefont
  {Z.}~\bibnamefont {Liu}}, \bibinfo {author} {\bibfnamefont {J.-P.}\
  \bibnamefont {Ma}}, \ and\ \bibinfo {author} {\bibfnamefont {J.-B.}\
  \bibnamefont {Zhang}} (\bibinfo {collaboration} {CLQCD}),\ }\href {\doibase
  10.1103/PhysRevD.87.014501} {\bibfield  {journal} {\bibinfo  {journal} {Phys.
  Rev.}\ }\textbf {\bibinfo {volume} {D87}},\ \bibinfo {pages} {014501}
  (\bibinfo {year} {2013})},\ \Eprint {http://arxiv.org/abs/1206.2086}
  {arXiv:1206.2086 [hep-lat]} \BibitemShut {NoStop}%
\bibitem [{\citenamefont {Yang}\ \emph {et~al.}(2012)\citenamefont {Yang},
  \citenamefont {Chen}, \citenamefont {Li},\ and\ \citenamefont
  {Liu}}]{Yang:2012gz}%
  \BibitemOpen
  \bibfield  {author} {\bibinfo {author} {\bibfnamefont {Y.-B.}\ \bibnamefont
  {Yang}}, \bibinfo {author} {\bibfnamefont {Y.}~\bibnamefont {Chen}}, \bibinfo
  {author} {\bibfnamefont {G.}~\bibnamefont {Li}}, \ and\ \bibinfo {author}
  {\bibfnamefont {K.-F.}\ \bibnamefont {Liu}},\ }\href {\doibase
  10.1103/PhysRevD.86.094511} {\bibfield  {journal} {\bibinfo  {journal} {Phys.
  Rev.}\ }\textbf {\bibinfo {volume} {D86}},\ \bibinfo {pages} {094511}
  (\bibinfo {year} {2012})},\ \Eprint {http://arxiv.org/abs/1202.2205}
  {arXiv:1202.2205 [hep-ph]} \BibitemShut {NoStop}%
\bibitem [{\citenamefont {Luscher}(1986)}]{Luscher:1986pf}%
  \BibitemOpen
  \bibfield  {author} {\bibinfo {author} {\bibfnamefont {M.}~\bibnamefont
  {Luscher}},\ }\href {\doibase 10.1007/BF01211097} {\bibfield  {journal}
  {\bibinfo  {journal} {Commun. Math. Phys.}\ }\textbf {\bibinfo {volume}
  {105}},\ \bibinfo {pages} {153} (\bibinfo {year} {1986})}\BibitemShut
  {NoStop}%
\bibitem [{\citenamefont {Luscher}(1991)}]{Luscher:1990ux}%
  \BibitemOpen
  \bibfield  {author} {\bibinfo {author} {\bibfnamefont {M.}~\bibnamefont
  {Luscher}},\ }\href {\doibase 10.1016/0550-3213(91)90366-6} {\bibfield
  {journal} {\bibinfo  {journal} {Nucl. Phys.}\ }\textbf {\bibinfo {volume}
  {B354}},\ \bibinfo {pages} {531} (\bibinfo {year} {1991})}\BibitemShut
  {NoStop}%
\bibitem [{\citenamefont {Aoki}\ \emph {et~al.}(2011)\citenamefont {Aoki} \emph
  {et~al.}}]{Aoki:2010dy}%
  \BibitemOpen
  \bibfield  {author} {\bibinfo {author} {\bibfnamefont {Y.}~\bibnamefont
  {Aoki}} \emph {et~al.} (\bibinfo {collaboration} {RBC, UKQCD}),\ }\href
  {\doibase 10.1103/PhysRevD.83.074508} {\bibfield  {journal} {\bibinfo
  {journal} {Phys. Rev.}\ }\textbf {\bibinfo {volume} {D83}},\ \bibinfo {pages}
  {074508} (\bibinfo {year} {2011})},\ \Eprint {http://arxiv.org/abs/1011.0892}
  {arXiv:1011.0892 [hep-lat]} \BibitemShut {NoStop}%
\bibitem [{\citenamefont {Mawhinney}(2019)}]{Mawhinney:2019cuc}%
  \BibitemOpen
  \bibfield  {author} {\bibinfo {author} {\bibfnamefont {R.~D.}\ \bibnamefont
  {Mawhinney}} (\bibinfo {collaboration} {RBC, UKQCD}),\ }\href@noop {} {\
  (\bibinfo {year} {2019})},\ \Eprint {http://arxiv.org/abs/1912.13150}
  {arXiv:1912.13150 [hep-lat]} \BibitemShut {NoStop}%
\bibitem [{\citenamefont {Chiu}\ and\ \citenamefont
  {Zenkin}(1999)}]{Chiu:1998gp}%
  \BibitemOpen
  \bibfield  {author} {\bibinfo {author} {\bibfnamefont {T.-W.}\ \bibnamefont
  {Chiu}}\ and\ \bibinfo {author} {\bibfnamefont {S.~V.}\ \bibnamefont
  {Zenkin}},\ }\href {\doibase 10.1103/PhysRevD.59.074501} {\bibfield
  {journal} {\bibinfo  {journal} {Phys. Rev.}\ }\textbf {\bibinfo {volume}
  {D59}},\ \bibinfo {pages} {074501} (\bibinfo {year} {1999})},\ \Eprint
  {http://arxiv.org/abs/hep-lat/9806019} {arXiv:hep-lat/9806019 [hep-lat]}
  \BibitemShut {NoStop}%
\bibitem [{\citenamefont {Zyla}\ \emph {et~al.}(2020)\citenamefont {Zyla} \emph
  {et~al.}}]{Zyla:2020zbs}%
  \BibitemOpen
  \bibfield  {author} {\bibinfo {author} {\bibfnamefont {P.~A.}\ \bibnamefont
  {Zyla}} \emph {et~al.} (\bibinfo {collaboration} {Particle Data Group}),\
  }\href {\doibase 10.1093/ptep/ptaa104} {\bibfield  {journal} {\bibinfo
  {journal} {PTEP}\ }\textbf {\bibinfo {volume} {2020}},\ \bibinfo {pages}
  {083C01} (\bibinfo {year} {2020})}\BibitemShut {NoStop}%
\bibitem [{\citenamefont {Liang}\ \emph {et~al.}(2014)\citenamefont {Liang},
  \citenamefont {Chen}, \citenamefont {Gong}, \citenamefont {Gui},
  \citenamefont {Liu}, \citenamefont {Liu},\ and\ \citenamefont
  {Yang}}]{Liang:2013eoa}%
  \BibitemOpen
  \bibfield  {author} {\bibinfo {author} {\bibfnamefont {J.}~\bibnamefont
  {Liang}}, \bibinfo {author} {\bibfnamefont {Y.}~\bibnamefont {Chen}},
  \bibinfo {author} {\bibfnamefont {M.}~\bibnamefont {Gong}}, \bibinfo {author}
  {\bibfnamefont {L.-C.}\ \bibnamefont {Gui}}, \bibinfo {author} {\bibfnamefont
  {K.-F.}\ \bibnamefont {Liu}}, \bibinfo {author} {\bibfnamefont
  {Z.}~\bibnamefont {Liu}}, \ and\ \bibinfo {author} {\bibfnamefont {Y.-B.}\
  \bibnamefont {Yang}},\ }\href {\doibase 10.1103/PhysRevD.89.094507}
  {\bibfield  {journal} {\bibinfo  {journal} {Phys. Rev. D}\ }\textbf {\bibinfo
  {volume} {89}},\ \bibinfo {pages} {094507} (\bibinfo {year} {2014})},\
  \Eprint {http://arxiv.org/abs/1310.3532} {arXiv:1310.3532 [hep-lat]}
  \BibitemShut {NoStop}%
\bibitem [{\citenamefont {Chang}\ \emph {et~al.}(2018)\citenamefont {Chang}
  \emph {et~al.}}]{Chang:2018uxx}%
  \BibitemOpen
  \bibfield  {author} {\bibinfo {author} {\bibfnamefont {C.~C.}\ \bibnamefont
  {Chang}} \emph {et~al.},\ }\href {\doibase 10.1038/s41586-018-0161-8}
  {\bibfield  {journal} {\bibinfo  {journal} {Nature}\ }\textbf {\bibinfo
  {volume} {558}},\ \bibinfo {pages} {91} (\bibinfo {year} {2018})},\ \Eprint
  {http://arxiv.org/abs/1805.12130} {arXiv:1805.12130 [hep-lat]} \BibitemShut
  {NoStop}%
\bibitem [{\citenamefont {Bhardwaj}\ \emph {et~al.}(2013)\citenamefont
  {Bhardwaj} \emph {et~al.}}]{Bhardwaj:2013rmw}%
  \BibitemOpen
  \bibfield  {author} {\bibinfo {author} {\bibfnamefont {V.}~\bibnamefont
  {Bhardwaj}} \emph {et~al.} (\bibinfo {collaboration} {Belle}),\ }\href
  {\doibase 10.1103/PhysRevLett.111.032001} {\bibfield  {journal} {\bibinfo
  {journal} {Phys. Rev. Lett.}\ }\textbf {\bibinfo {volume} {111}},\ \bibinfo
  {pages} {032001} (\bibinfo {year} {2013})},\ \Eprint
  {http://arxiv.org/abs/1304.3975} {arXiv:1304.3975 [hep-ex]} \BibitemShut
  {NoStop}%
\bibitem [{\citenamefont {Ablikim}\ \emph {et~al.}(2015)\citenamefont {Ablikim}
  \emph {et~al.}}]{Ablikim:2015dlj}%
  \BibitemOpen
  \bibfield  {author} {\bibinfo {author} {\bibfnamefont {M.}~\bibnamefont
  {Ablikim}} \emph {et~al.} (\bibinfo {collaboration} {BESIII}),\ }\href
  {\doibase 10.1103/PhysRevLett.115.011803} {\bibfield  {journal} {\bibinfo
  {journal} {Phys. Rev. Lett.}\ }\textbf {\bibinfo {volume} {115}},\ \bibinfo
  {pages} {011803} (\bibinfo {year} {2015})},\ \Eprint
  {http://arxiv.org/abs/1503.08203} {arXiv:1503.08203 [hep-ex]} \BibitemShut
  {NoStop}%
\bibitem [{\citenamefont {Aaij}\ \emph {et~al.}(2019)\citenamefont {Aaij} \emph
  {et~al.}}]{Aaij:2019evc}%
  \BibitemOpen
  \bibfield  {author} {\bibinfo {author} {\bibfnamefont {R.}~\bibnamefont
  {Aaij}} \emph {et~al.} (\bibinfo {collaboration} {LHCb}),\ }\href {\doibase
  10.1007/JHEP07(2019)035} {\bibfield  {journal} {\bibinfo  {journal} {JHEP}\
  }\textbf {\bibinfo {volume} {07}},\ \bibinfo {pages} {035} (\bibinfo {year}
  {2019})},\ \Eprint {http://arxiv.org/abs/1903.12240} {arXiv:1903.12240
  [hep-ex]} \BibitemShut {NoStop}%
\bibitem [{\citenamefont {Maezawa}\ and\ \citenamefont
  {Petreczky}(2016)}]{Maezawa:2016vgv}%
  \BibitemOpen
  \bibfield  {author} {\bibinfo {author} {\bibfnamefont {Y.}~\bibnamefont
  {Maezawa}}\ and\ \bibinfo {author} {\bibfnamefont {P.}~\bibnamefont
  {Petreczky}},\ }\href {\doibase 10.1103/PhysRevD.94.034507} {\bibfield
  {journal} {\bibinfo  {journal} {Phys. Rev. D}\ }\textbf {\bibinfo {volume}
  {94}},\ \bibinfo {pages} {034507} (\bibinfo {year} {2016})},\ \Eprint
  {http://arxiv.org/abs/1606.08798} {arXiv:1606.08798 [hep-lat]} \BibitemShut
  {NoStop}%
\bibitem [{\citenamefont {Yang}\ \emph
  {et~al.}(2015{\natexlab{b}})\citenamefont {Yang} \emph
  {et~al.}}]{Yang:2014sea}%
  \BibitemOpen
  \bibfield  {author} {\bibinfo {author} {\bibfnamefont {Y.-B.}\ \bibnamefont
  {Yang}} \emph {et~al.},\ }\href {\doibase 10.1103/PhysRevD.92.034517}
  {\bibfield  {journal} {\bibinfo  {journal} {Phys. Rev.}\ }\textbf {\bibinfo
  {volume} {D92}},\ \bibinfo {pages} {034517} (\bibinfo {year}
  {2015}{\natexlab{b}})},\ \Eprint {http://arxiv.org/abs/1410.3343}
  {arXiv:1410.3343 [hep-lat]} \BibitemShut {NoStop}%
\bibitem [{\citenamefont {Alexandrou}\ \emph {et~al.}(2020)\citenamefont
  {Alexandrou}, \citenamefont {Bacchio}, \citenamefont {Constantinou},
  \citenamefont {Finkenrath}, \citenamefont {Hadjiyiannakou}, \citenamefont
  {Jansen}, \citenamefont {Koutsou}, \citenamefont {Panagopoulos},\ and\
  \citenamefont {Spanoudes}}]{Alexandrou:2020sml}%
  \BibitemOpen
  \bibfield  {author} {\bibinfo {author} {\bibfnamefont {C.}~\bibnamefont
  {Alexandrou}}, \bibinfo {author} {\bibfnamefont {S.}~\bibnamefont {Bacchio}},
  \bibinfo {author} {\bibfnamefont {M.}~\bibnamefont {Constantinou}}, \bibinfo
  {author} {\bibfnamefont {J.}~\bibnamefont {Finkenrath}}, \bibinfo {author}
  {\bibfnamefont {K.}~\bibnamefont {Hadjiyiannakou}}, \bibinfo {author}
  {\bibfnamefont {K.}~\bibnamefont {Jansen}}, \bibinfo {author} {\bibfnamefont
  {G.}~\bibnamefont {Koutsou}}, \bibinfo {author} {\bibfnamefont
  {H.}~\bibnamefont {Panagopoulos}}, \ and\ \bibinfo {author} {\bibfnamefont
  {G.}~\bibnamefont {Spanoudes}},\ }\href {\doibase
  10.1103/PhysRevD.101.094513} {\bibfield  {journal} {\bibinfo  {journal}
  {Phys. Rev.}\ }\textbf {\bibinfo {volume} {D101}},\ \bibinfo {pages} {094513}
  (\bibinfo {year} {2020})},\ \Eprint {http://arxiv.org/abs/2003.08486}
  {arXiv:2003.08486 [hep-lat]} \BibitemShut {NoStop}%
\bibitem [{\citenamefont {Alexandru}\ \emph {et~al.}(2012)\citenamefont
  {Alexandru}, \citenamefont {Pelissier}, \citenamefont {Gamari},\ and\
  \citenamefont {Lee}}]{Alexandru:2011ee}%
  \BibitemOpen
  \bibfield  {author} {\bibinfo {author} {\bibfnamefont {A.}~\bibnamefont
  {Alexandru}}, \bibinfo {author} {\bibfnamefont {C.}~\bibnamefont
  {Pelissier}}, \bibinfo {author} {\bibfnamefont {B.}~\bibnamefont {Gamari}}, \
  and\ \bibinfo {author} {\bibfnamefont {F.}~\bibnamefont {Lee}},\ }\href
  {\doibase 10.1016/j.jcp.2011.11.003} {\bibfield  {journal} {\bibinfo
  {journal} {J. Comput. Phys.}\ }\textbf {\bibinfo {volume} {231}},\ \bibinfo
  {pages} {1866} (\bibinfo {year} {2012})},\ \Eprint
  {http://arxiv.org/abs/1103.5103} {arXiv:1103.5103 [hep-lat]} \BibitemShut
  {NoStop}%
\bibitem [{\citenamefont {Alexandru}\ \emph {et~al.}(2011)\citenamefont
  {Alexandru}, \citenamefont {Lujan}, \citenamefont {Pelissier}, \citenamefont
  {Gamari},\ and\ \citenamefont {Lee}}]{Alexandru:2011sc}%
  \BibitemOpen
  \bibfield  {author} {\bibinfo {author} {\bibfnamefont {A.}~\bibnamefont
  {Alexandru}}, \bibinfo {author} {\bibfnamefont {M.}~\bibnamefont {Lujan}},
  \bibinfo {author} {\bibfnamefont {C.}~\bibnamefont {Pelissier}}, \bibinfo
  {author} {\bibfnamefont {B.}~\bibnamefont {Gamari}}, \ and\ \bibinfo {author}
  {\bibfnamefont {F.~X.}\ \bibnamefont {Lee}}\ }(\bibinfo {year} {2011})\ pp.\
  \bibinfo {pages} {123--130},\ \Eprint {http://arxiv.org/abs/1106.4964}
  {arXiv:1106.4964 [hep-lat]} \BibitemShut {NoStop}%
\bibitem [{\citenamefont {Bi}\ \emph {et~al.}(2020)\citenamefont {Bi},
  \citenamefont {Xiao}, \citenamefont {Guo}, \citenamefont {Gong},
  \citenamefont {Sun}, \citenamefont {Xu},\ and\ \citenamefont
  {Yang}}]{Bi:2020wpt}%
  \BibitemOpen
  \bibfield  {author} {\bibinfo {author} {\bibfnamefont {Y.-J.}\ \bibnamefont
  {Bi}}, \bibinfo {author} {\bibfnamefont {Y.}~\bibnamefont {Xiao}}, \bibinfo
  {author} {\bibfnamefont {W.-Y.}\ \bibnamefont {Guo}}, \bibinfo {author}
  {\bibfnamefont {M.}~\bibnamefont {Gong}}, \bibinfo {author} {\bibfnamefont
  {P.}~\bibnamefont {Sun}}, \bibinfo {author} {\bibfnamefont {S.}~\bibnamefont
  {Xu}}, \ and\ \bibinfo {author} {\bibfnamefont {Y.-B.}\ \bibnamefont
  {Yang}},\ }\href {\doibase 10.22323/1.363.0286} {\bibfield  {journal}
  {\bibinfo  {journal} {PoS}\ }\textbf {\bibinfo {volume} {LATTICE2019}},\
  \bibinfo {pages} {286} (\bibinfo {year} {2020})},\ \Eprint
  {http://arxiv.org/abs/2001.05706} {arXiv:2001.05706 [hep-lat]} \BibitemShut
  {NoStop}%
\end{thebibliography}%

\end{document}